\documentclass[11pt]{article}
\usepackage{xspace}
\usepackage{graphicx}
\usepackage{amsmath}
\usepackage{amssymb}
\usepackage{color}

\definecolor{Red}{rgb}{1,0,0}
\definecolor{Green}{rgb}{0,1,0}
\definecolor{Blue}{rgb}{0,0,1}
\definecolor{Black}{rgb}{0,0,0}

\def\beq{\begin{equation}}
\def\eeq#1{\label{#1}\end{equation}}
\def\eeqn{\end{equation}}

\def\beqa{\begin{eqnarray}}
\def\eeqa#1{\label{#1}\end{eqnarray}}
\def\eeqan{\end{eqnarray}}

\let\bar=\overbar

\def\L{{\cal L}}

\def\hc{{\mbox{\rm h.c.}}}
\def\tr{{\mbox{\rm tr}}}

\def\Dslash{\not{\hbox{\kern-4pt $D$}}}
\def\dslash{\not{\hbox{\kern-2pt $\del$}}}

\def\ee{e^+e^-}

\def\msb{{\bar{\ssstyle M \kern -1pt S}}}

\usepackage{mathrsfs}
\usepackage{bbm}
\usepackage{cite}

\newcommand{\LL}{\mathscr{L}}

\def\cY{{\bf Y}}
\def\cy{{\bf y}}

\def\be{\begin{equation}}
\def\ee{\end{equation}}
\def\beq{\begin{equation}}
\def\eeq{\end{equation}}
\def\bc{\begin{center}}
\def\ec{\end{center}}
\def\bea{\begin{eqnarray}}
\def\eea{\end{eqnarray}}

\newcommand{\mean}[1]{\langle#1\rangle}

\newcommand{\ov}[1]{\overline{#1}}
\newcommand{\unity}{\mathbbm{1}}
\newcommand{\ep}{\epsilon}
\newcommand{\diag}{\mathrm{diag}}

\def\T2K{{\sc T2K }}

\def\L{{\rm L}}
\def\R{{\rm R}}

\newcommand{\refcite}[1]{Ref.~\cite{#1}}

\def\Title#1{\begin{center} {\Large {\bf #1} } \end{center}}

\begin{document}

\Title{On Continuous Flavour Symmetries for Neutrinos}

\bigskip\bigskip


\begin{raggedright}  

{\it Luca Merlo\index{Merlo, L.},\\
Departamento de F\'isica Te\'orica and Instituto de F\'{\i}sica Te\'orica, IFT-UAM/CSIC,\\
Universidad Aut\'onoma de Madrid, Cantoblanco, 28049, Madrid, Spain}\\

\end{raggedright}
\vspace{1.cm}

{\small
\begin{flushleft}
\emph{To appear in the proceedings of the Prospects in Neutrino Physics Conference, 15 -- 17 December, 2014, held at Queen Mary University of London, UK.}
\end{flushleft}
}

%
%
\section{Introduction}

Before the discovery of a non-vanishing reactor angle, discrete symmetries were deeply implemented in the construction of flavour models to explain the flavour puzzle. In particular, it was a common feature of this class of models the prediction in first approximation of the PMNS matrix with a vanishing reactor angle and a maximal atmospheric one (e.g. Ref.~\cite{Altarelli:2010gt}). 

With a sizable reactor angle~\cite{GonzalezGarcia:2012sz}, these models underwent to a severe loss of attractiveness. To achieve a model in agreement with the new data, a few strategies have been followed~\cite{Luhntalk}: introduction of additional parameters in preexisting minimal models; implementation of features that allow next order corrections only in specific directions in the flavour space; search for alternative mixing patterns or flavour symmetries that lead already in first approximation to $\theta_{13}\neq 0^\circ$ (see for example Refs.~\cite{Altarelli:2009gn} and references therein). In summary, the latest neutrino data can still be described in the context of discrete symmetries, but at the prize of fine-tunings and/or eccentric mechanisms. 

Sum rules among neutrino masses and mixing angles are usually present in these models and are useful as tests at experiments~\cite{Ballett:2013wya}. Furthermore, studies on flavour violating observables~\cite{Feruglio:2008ht}, on the connection with astroparticle physics~\cite{Bertuzzo:2009im}, on the parameter running~\cite{Dighe:2006sr} and on the role of the CP symmetry~\cite{Feruglio:2012cw} have been performed to fully workout this framework. On the other side, the scalar and messenger sectors of these models are in general very complicated~\cite{Altarelli:2005yp}, it is not easy to provide a successful description of the quark sector~\cite{Feruglio:2007uu}, and the selection of a specific discrete symmetry usually does not follow from a more general criterium~\cite{Altarelli:2006kg}, but it is just a matter of taste.

Even if it is still worth to search for a realistic model based on discrete symmetries, the many drawbacks suggest to investigate alternative approaches: here the focus will be on continuous symmetries such as the simplest Abelian $U(1)$ or non-Abelian groups.

%
%
\boldmath
\section{Abelian models}
\unboldmath
\label{u1mod}

Models based on the Abelian $U(1)$ group are sometimes preferred with respect to those based on discrete symmetries for a series of reasons. First of all, the Abelian $U(1)$ group is an element already present in the Standard Model (SM) and in many beyond SM (BSM) theories. Furthermore, it has been shown much time ago that the quark sector~\cite{Froggatt:1978nt} is easily described in this context. In addition, the formulation of a model based on the $U(1)$ symmetry, in the supersymmetric context as the holomorphicity of the superpotential simplifies the construction of the Yukawa interactions, is simple and elegant:\\
- The flavour symmetry acts horizontally on leptons and the charges can be written as $e^c\sim(n_1^\R,n_2^\R,0)$, with $n_1^\R>n_2^\R>0$, for the $SU(2)_\L$ lepton singlets and as $\ell\sim(n_1^\L,n_2^\L,0)$ for the $SU(2)_\L$ lepton doublets. The Higgs fields are taken to be $U(1)$-singlets.\\
- Once leptons have $U(1)$ charges, the Yukawa terms are no longer invariant under the action of the flavour symmetry. To formally recover the invariance, a new scalar field, the flavon $\theta$, can be introduced that transforms non-trivially only under $U(1)$, with charge $n_\theta$. Then, the Yukawa Lagrangian can be written as 
\beq
-\mathcal{L}_Y =\,(y_e)_{ij}\,\ell_i\,H_d\, e^c_j \left(\dfrac{\theta}{\Lambda}\right)^{p_e}+(y_\nu)_{ij}\,\dfrac{\ell_i\ell_j H_u H_u}{\Lambda_\L}\left(\dfrac{\theta}{\Lambda}\right)^{p_\nu} +\text{h.c.}
\label{Yukawas}
\eeq
where $\Lambda$ is the cut-off of the effective flavour theory and $\Lambda_\L$ the scale of the lepton number violation. $(y_e)_{ij}$ and $(y_\nu)_{ij}$ are free parameters, taken to be complex and with modulus of order 1. $p_e$ and $p_\nu$ are suitable powers of the dimensionless ratio $\theta/\Lambda$ necessary to enforce the invariance under the flavour symmetry. 
Without loss of generality, we can fix $n_\theta=-1$; consequently,  $n_1,n_2>0$ to assure that the Lagrangian expansion is well defined. 
Here, neutrino masses are described by the effective Weinberg operator (see i.e. Refs.~\cite{Altarelli:2000fu,Altarelli:2012ia}).\\
- Once the flavon and the Higgs fields develop non-vanishing vacuum expectation values (VEVs), the flavour and electroweak symmetries are broken and mass matrices arise from the Yukawa Lagrangian. Defining, $\ep\equiv\mean{\theta}/\Lambda<1$, a useful parametrisation for the Yukawa matrices is given by 
$Y_e=F_{e^c}\,y_e\,F_{\ell}$ and $Y_\nu=F_{\ell}\,y_\nu\,F_{\ell}$, where $F_f=\diag(\ep^{n_{f1}},\ep^{n_{f2}},\ep^{n_{f3}})$. Following Ref.~\cite{Altarelli:2012ia,Bergstrom:2014owa}, the charges will be taken to be integers.

\boldmath
\subsection{Specific $U(1)$ models}
\label{sec:SpecificU1}
\unboldmath

In the following, focussing on constructions where neutrino masses are described by the Weinberg operator, two specific models will be considered: the model $A$ representative of anarchical constructions and the model $H$ representative of hierarchical ones,
\begin{center}
\begin{tabular}{ccc}
{Anarchy ($A$)}:& $e_\R\sim(3,1,0)$\,, & $\ell_\L\sim(0,0,0)$\,, \\
{Hierarchy ($H$)}:& $e_\R\sim(8,3,0)$\,, & $\ell_\L\sim(2,1,0)$\,.
\end{tabular}
\end{center}
$A$ encodes the idea that an even structureless mass matrix can lead to a correct description of neutrino data: this mass matrix is characterised by entries that are random numbers which, under the additional requirement of basis  invariance, leads to a unique measure of the mixing matrix -- the Haar measure~\cite{Hall:1999sn,deGouvea:2012ac,Lu:2014cla}. It has been claimed that such matrix generically prefers large mixings~\cite{Hall:1999sn} and that the observed sizable deviation from a zero reactor angle seems to favour anarchical models when compared to other more symmetric constructions \cite{deGouvea:2012ac}. However, as discussed in \refcite{Espinosa:2003qz}, how much a large value of a parameter is preferred can depend strongly on the definition of ``preferred'' and of ``large''.

It has been suggested in Ref.~\cite{Altarelli:2012ia,Bergstrom:2014owa} (see Ref.~\cite{Hirsch:2001he} for an earlier study) that the performances of anarchical models, formulated in a $U(1)$ context giving no charges to the left-handed fields, in reproducing the 2012 neutrino data are worse than those of models constructed upon the $U(1)$ flavour symmetry. In the latter ones, the small neutrino parameters are due to the built-in hierarchies and not due to chance. The construction $H$ considered in Ref.~\cite{Bergstrom:2014owa} has been shown by mean of the Bayesian inference to be the best one to describe the data, among all the possible $U(1)$ models. 

The textures for the charged leptons $Y_e$ and neutrino $Y_\nu$ Yukawa matrices are as follows:
\beq
\begin{aligned}
A:\quad 
Y_e&=\left(  \begin{matrix}  \ep^3 &    \ep &   1 \\  \ep^3 &    \ep &   1 \\ \ep^3 &    \ep &   1 \end{matrix}\right)\,,\;
Y_\nu= \left(  \begin{matrix}  1 &    1 &   1 \\  1 &   1 &  1 \\ 1 &   1 &   1 \end{matrix}\right)\,,\\
H:\quad 
Y_e&= \left(  \begin{matrix}  \ep^{10} &    \ep^6 &   \ep^2 \\  \ep^9 &    \ep^5 &   \ep \\ \ep^8 &   \ep^4 &   1 \end{matrix} \right) ,\;
Y_\nu= \left(  \begin{matrix}  \ep^4 &    \ep^3 &   \ep^2 \\  \ep^3 &   \ep^2 &  \ep \\ \ep^2 &   \ep &   1 \end{matrix}     \right)\,.
\end{aligned}
\label{NewModels} 
\eeq
In the spirit of $U(1)$ models, the coefficients in front of $\ep^n$ are expected to be complex numbers with absolute values of ${\cal O}(1)$ and arbitrary phases. As $Y_\nu$ is a symmetric matrix, the total number of parameters that should be considered in the analysis is $30$, from the Yukawa matrices, plus the unknown value of $\ep$.

While for the details of the analysis we refer to the original publication \refcite{Bergstrom:2014owa}, here we just comment on the results of the comparison between $A$ and $H$. Both the models have a $\chi^2$-minimum of zero and therefore a $\chi^2$-analysis can never exclude any of the models or be meaningful to compare them. On the other side, a Bayesian analysis allows instead a quantitative comparison of the models: the ratio between the logarithms of the evidences of $H$ normalised to the evidence of $A$, i.e., the Bayes factor between $H$ and $A$, is $\log B \simeq 3\div4.5$, depending on the prior on $\ep$. The uncertainty on the logarithms of the Bayes factors is about $0.2$. Accordingly to the Jeffreys scale, these values translate in a moderate evidence in favour of $H$ with respect to $A$.

Regarding the data adopted here, improved measurements of the oscillation parameters cannot further discriminate between the models. Instead, there are other observables which could be accurately measured in future experiments, and in principle could be used to distinguish between the models. These are primarily the CP-phase $\delta$ and observables related to the values of neutrino masses ($m_ {ee}$, $m_\beta$, $\Sigma$). See \refcite{Bergstrom:2014owa} for details.

\boldmath
\section{Non-Abelian models}
\unboldmath
\label{MLFVmod}

One of the main problematics of dealing with Abelian symmetries is the fact that the three fermion generations are independent from each other, translating in the large number of free parameters. On the other hand, in the context of non-Abelian symmetries, when fermions transform with multidimensional representations, the three families are connected one to the others, reducing the number of free parameters and therefore increasing predictivity. 

Non-Abelian continuous symmetries have also been deeply investigated in the flavour sector, but mainly connected to the Minimal Flavour Violation (MFV)~\cite{Chivukula:1987py} ansatz: i.e. the requirement that 
all sources of flavour violation in the SM and BSM are described at low-energies uniquely in terms 
of the known fermion masses and mixings. Several distinct models formulated in this framework~\cite{D'Ambrosio:2002ex,Cirigliano:2005ck,Gavela:2009cd,Alonso:2011jd,Davidson:2006bd,Alonso:2011yg,Alonso:2012fy,Alonso:2013mca,Alonso:2013nca} turn out to be consistent with TeV new physics. 

The power of MFV descends from the fact that it exploits the symmetries that the SM itself contains in a 
certain limit: that of massless fermions. For example, in the case of the Type I Seesaw mechanism with three 
RH neutrinos added to the SM spectrum, the flavour symmetry of the full Lagrangian, when Yukawa couplings and 
the RH neutrino masses are set to zero, is $G_f=G^q_f\times G^\ell_f$ with $G^q_f=U(3)_{Q_L}\times U(3)_{U_R}\times U(3)_{D_R}$ and $G^\ell_f=U(3)_{\ell_L}\times U(3)_{E_R}\times U(3)_{N}$.
Under the flavour symmetry group $G_f$ fermion fields transform as
\beq
\begin{gathered}
Q_L\sim(3,1,1)_{G_f^q}\,,\qquad
U_R\sim(1,3,1)_{G_f^q}\,,\qquad
D_R\sim(1,1,3)_{G_f^q}\,,\\
\ell_L\sim(3,1,1)_{G_f^\ell}\,,\qquad
E_R\sim(1,3,1)_{G_f^\ell}\,,\qquad
N_R\sim(1,1,3)_{G_f^\ell}\,.
\end{gathered}
\label{FermionTransf}
\eeq
The Yukawa Lagrangian for the Type I Seesaw mechanism, then, reads:
\beq
-\LL_Y=\ov{Q}_LY_DHD_R+\ov{Q}_LY_U\tilde{H}U_R+\ov{\ell}_LY_EHE_R+\ov{\ell}_LY_\nu\tilde{H}N_R+\ov{N}^c_R\dfrac{M_N}{2}N_R+\hc
\label{Lagrangian}
\eeq 
To introduce $\LL_Y$ without explicitly breaking $G_f$, the Yukawa matrices $Y_i$ and the mass matrix for the 
RH neutrinos $M_N$ have to be promoted to be spurion fields transforming under the flavour symmetry as:
\beq
Y_U\sim(3,\bar3,1)_{G_f^q},\,
Y_D\sim(3,1,\bar3)_{G_f^q},\,
Y_E\sim(3,\bar3,1)_{G_f^\ell},\,
Y_\nu\sim(3,1,\bar3)_{G_f^\ell},\,
M_N\sim (1,1,\bar6)_{G_f^\ell}.
\label{YTransf}
\eeq
The quark masses and mixings are correctly reproduced once the quark spurion Yukawas get background values as
$Y_U =V^\dag\,\cy_U$ and $Y_D =\cy_D$ where $\cy_{U,D}$ are diagonal matrices with Yukawa eigenvalues as diagonal entries, and $V$ a unitary matrix 
that in good approximation coincides with the CKM matrix. 

When discussing the MFV ansatz in the leptonic sector one has at disposal three different spurions, as can be 
evinced from the list in Eq.~(\ref{YTransf}). The number of parameters that can be introduced in the model through 
these spurions is much larger than the low energy observables. This in general prevents a direct link among neutrino 
parameters and FV observables. The usual way adopted in the literature to lower the number of parameters consists 
in reducing the number of spurions from three to two: for example Ref.~\cite{Cirigliano:2005ck} takes $M_N\propto\unity$; in Ref.~\cite{Gavela:2009cd}, a two-family RH neutrino model is considered with $M_N\propto\sigma_1$; 
finally in Ref.~\cite{Alonso:2011jd} $Y^\dag_\nu Y_\nu\propto\unity$ is assumed. 

An unifying description for all these models can be obtained by introducing the Casas-Ibarra parametrization: in the basis of diagonal mass matrices for RH neutrinos, LH neutrinos and charged 
leptons, the neutrino Yukawa coupling can be written as 
\beq
Y_\nu=\dfrac{1}{v}U\sqrt{\hat m_\nu}R\sqrt{\hat M_N},
\label{CIpar}
\eeq
where $v$ is the electroweak vev, the hatted matrices are light and heavy neutrino diagonal mass matrices, $U$ refers 
to the PMNS mixing matrix and $R$ is a complex orthogonal matrix, $R^T R=\unity$. A correct description of lepton 
masses and mixings is achieved assuming that $Y_E$ acquires a background value parametrised by a diagonal matrix,
$Y_E=\cy_E\equiv{\rm diag}(y_e,\,y_\mu,\,y_\tau)$, while the remaining spurion, $M_N$ or $Y_\nu$, accounts for the neutrino masses and the PMNS matrix (see 
Refs.~\cite{Cirigliano:2005ck,Gavela:2009cd,Alonso:2011jd}).

\boldmath
\subsection{Dynamical Yukawas}
\label{Sect:DynamicalYukawas}
\unboldmath

Despite of the phenomenological success, it has to be noticed that, however, MFV does not provide by itself 
any explanation of the origin of fermion masses and/or mixing, or equivalently does not provide any explanation 
for the background values of the Yukawa spurions. This observation motivates the studies performed in 
Refs.~\cite{Alonso:2011yg,Alonso:2012fy,Alonso:2013mca,Alonso:2013nca}, where the Yukawa spurions are promoted to dynamical scalar fields: the case in which a one-to-one correlation among Yukawa couplings and fields is assumed, $Y_i\equiv \mean{\cY_i}/\Lambda_f$, is discussed at length. The scalar potential constructed out of these fields was studied in Refs.~\cite{Alonso:2011yg,Alonso:2012fy,Alonso:2013mca,Alonso:2013nca}, considering renormalisable operators (and adding also lower-order non-renormalisable terms for the quark case): these effective Lagrangian expansions are possible under the assumption that the ratio of the flavon vevs and the cutoff scale of the theory is smaller than 1, condition that is always satisfied but for the top Yukawa coupling. In this case a non-linear description would be more suitable. We focus here only on the lepton sector, while for the quark sector we refer to the original article in \refcite{Alonso:2011yg}.
\\

\noindent{\bf - Two-family case -}

It is instructive and interesting to start with a toy model with only two generations~\cite{Alonso:2012fy}. Under the assumption of degenerate RH neutrino masses, $M_1=M_2\equiv M$, only the spurion fields $Y_E$ and $Y_\nu$ are promoted to dynamical fields, $\cY_E$ and $\cY_\nu$, and the flavour symmetry in this case is $G_f^\ell=U(2)_{\ell_L}\times U(2)_{E_R}\times O(2)_{N}$.
Only five independent invariants can be obtained at the renormalisable level:
\beq
\begin{gathered}
\tr\left[\cY_E\cY_E^\dagger\right],\,\,\,\,\,
\tr\left[\cY_\nu \cY_\nu^\dagger\right],\,\, \,\,\,
\tr\left[\left(\cY_E\cY_E^\dagger\right)^2\right],\,\,\,\,\,
\tr\left[\left(\cY_\nu \cY_\nu^\dagger\right)^2\right],\,\,\,\,\,
\tr\left[\left(\cY_\nu\sigma_2\cY_\nu^\dagger\right)^2\right].
\end{gathered}
\label{LeptonInvariants}
\eeq
All the terms account for the lepton masses, but the last one that fixes the mixing angle. By adopting the Casas-Ibarra parametrisation in Eq.~(\ref{CIpar}) and minimising the scalar potential with respect to the angle $\theta$ and the Majorana phase $\alpha$, the following two conditions result:
\beq
(y^2-y'^2)\sqrt{m_{\nu_2}m_{\nu_1}}\sin2\theta\cos2\alpha=0\,,\qquad\qquad
\tan2\theta=\sin2\alpha\frac{y^2-y'^2}{y^2+y'^2}
\frac{2\sqrt{m_{\nu_2}m_{\nu_1}}}{m_{\nu_2}-m_{\nu_1}}\,,
\eeq
where $y$ and $y'$ are two parameters of $Y_\nu$ (see Ref.~\cite{Alonso:2012fy} for details). The first condition implies a maximal Majorana phase, $\alpha=\pi/4$ or $\alpha=3\pi/4$, for a non-trivial mixing angle. However, this does not imply observability of CP violation at experiments, 
as the relative Majorana phase among the two neutrino eigenvalues is $\pi/2$. The second condition above represents a link among the size of mixing angle and the type of the neutrino spectrum: a large mixing angle is obtained from almost degenerate masses, while a small angle follows in the hierarchical case. 
\\

\noindent{\bf - Generalisation to the three-family case -}

Moving to the realistic scenario of three families, but still considering $G_f^\ell$ containing a $O(2)_{N}$ factor, $N_R$ ($N'_R$) is a doublet (singlet) of $O(2)_{N}$. Correspondingly, the neutrino Yukawa field accounts for two components: a doublet and a singlet of $O(2)_{N}$, $\cY_\nu\sim(3,1,\bar2)$ and $\cY'_\nu\sim(3,1,1)$.
The leptonic flavour Lagrangian is given in this case by
\beq
-\LL_Y=\ov{\ell}_LY_EHE_R+\ov{\ell}_LY'_\nu\tilde{H}N'_R+\ov{\ell}_LY_\nu\tilde{H}N_R+\dfrac{M'}{2}\ov{N}^{\prime c}_RN'_R+\dfrac{M}{2}\ov{N}^c_R\unity N_R+\hc\,.
\label{Lagrangian2}
\eeq 
Once the Yukawa flavons develop vevs, the light neutrino mass matrix is generated: 
\beq
m_\nu=\dfrac{v^2}{M'}Y'_\nu Y^{\prime T}_\nu+\dfrac{v^2}{M}Y_\nu Y^T_\nu\,.
\eeq

A total of nine independent invariants at the renormalisable level can be constructed in this case, namely
\beq
\begin{gathered}
\tr\left[\cY_E\cY_E^\dagger\right]\,,\quad
\tr\left[\cY_\nu\cY_\nu^\dagger\right]\,,\quad
\cY^{\prime \dag}_\nu \cY'_\nu\,,\quad
\tr\left[\left(\cY_E\cY_E^\dagger\right)^2\right]\,, \quad 
\tr\left[\left(\cY_\nu\cY_\nu^\dagger\right)^2\right]\,,\\
\tr\left[\cY_E\cY_E^\dagger\cY_\nu\cY_\nu^\dagger\right]\,, \qquad 
\tr\left[\cY_\nu\cY_\nu^T\cY_\nu^*\cY_\nu^\dag\right]\,,\qquad
\cY^{\prime \dag}_\nu \cY_E \cY^\dag_E \cY'_\nu \,,\qquad
\cY^{\prime \dag}_\nu \cY_\nu \cY^\dag_\nu \cY'_\nu \,.
\end{gathered}
\eeq
When considering the minimisation of the scalar potential, there are four analytical solutions: 
\beq
\begin{aligned}
1)\begin{cases}
\tan2\theta_{12}=z/z'\\
m_{\nu_1}\neq m_{\nu_2}\\
m_{\nu_3}=0
\end{cases}
2)
\begin{cases}
\theta_{12}=\pi/4\\
m_{\nu_1}= m_{\nu_2}\neq m_{\nu_3}\\
\alpha=\pi/4
\end{cases}
3)\begin{cases}
\theta_{23}=\pi/4\\
m_{\nu_1}\neq m_{\nu_2} = m_{\nu_3}\\
\alpha=\pi/4
\end{cases}
4)\begin{cases}
\tan2\theta_{23}=z/z'\\
m_{\nu_2}\neq m_{\nu_3}\\
m_{\nu_1}=0
\end{cases}
\end{aligned}
\label{configurations}
\eeq

Case 1 (4) describes an inverse (direct) hierarchical spectrum and only one sizable mixing angle, the solar (atmospheric) one. In case 2, the light neutrinos ${\nu_1}$ and ${\nu_2}$ are degenerate  and both mass orderings (hierarchical or degenerate) can be accommodated, while a maximal solar angle is predicted. Finally, case 3 corresponds to degenerate $\nu_2$ and $\nu_3$: a realistic scenario points to three almost degenerate neutrinos. Note that cases 2 and 3 encompass two degenerate neutrinos and  the relative Majorana phase between the two degenerate states is $\pi/2$. 

Cases 1-4 only account for one sizable angle. Realistic configurations with three non-trivial angles, however, follow in a straightforward way when interpolating between these four cases, at the prize of non-exact solutions that depend on the parameters of the of the scalar potential. The setup appears very promising, though, as all three angles can be naturally non-vanishing and moreover the number of free parameters is smaller than the number of observables, leading to predictive scenarios in which mixing angles and Majorana phases are linked to the spectrum.\\

It is finally interesting to consider the case with three degenerate RH neutrinos\cite{Alonso:2013mca,Alonso:2013nca}. The flavour symmetry is $G_f=U(3)_{\ell_L}\times U(3)_{E_R} \times O(3)_{N}$ and the basis of invariants is composed of the operators in Eq.~(\ref{LeptonInvariants}). The study of the extrema of these invariants has been presented in Ref.~\cite{Alonso:2013nca} and from the minimisation of the potential it follows that one of the possible configurations is 3) in Eq.~(\ref{configurations}). In the normal or inverse hierarchical case, two of the light neutrinos are degenerate in mass and a maximal angle and a maximal Majorana phase arise in their corresponding sector. On the other hand, if the third light neutrino is almost degenerate with the other two, then the perturbations split the spectrum and a second sizable angle arises~\cite{Alonso:2013nca}.

In summary, these results indicate that a realistic solution for the Flavour Puzzle in the lepton sector requires three RH neutrinos, two of which must be degenerate. All three light neutrinos would therefore acquire masses, and the precise values of the mixing angles and Majorana phases are related to the specific light mass spectrum, result that is almost exclusively a feature of continuous non-Abelian symmetries.

\bigskip
\section{Acknowledgments}

I acknowledge partial support by European Union FP7 ITN INVISIBLES (Marie Curie Actions, PITN-GA-2011-289442), by the Juan de la Cierva programme (JCI-2011-09244) and by the Spanish MINECOs ``Centro de Excelencia Severo Ochoa'' Programme under grant SEV-2012-0249. Finally, I thank the organisers and the conveners of the ``NuPhys2014: Prospects in Neutrino Physics'' conference for the kind invitation and for their efforts in organising this enjoyable meeting.

\providecommand{\href}[2]{#2}\begingroup\raggedright\endgroup

\end{document}